\documentclass[fleqn,twoside]{article}%
\usepackage{amsmath}
\usepackage{amsfonts}
\usepackage{amssymb}
\usepackage{graphicx}
\usepackage{bm}

\newcommand{\be}{\begin{equation}}
\newcommand{\ee}{\end{equation}}
\newcommand{\bey}{\begin{eqnarray}}
\newcommand{\eey}{\end{eqnarray}}
\def\bes{\begin{equation}\begin{split}&}
\newcommand{\bi}{\bibitem}

\title  {Noether symmetry in $f(T)$ teleparallel gravity.}
\author{Nayem Sk$^{1}$.\\
~~~~~~~~~~~~~~~\\
$^1$Dept. of Physics, University of Kalyani, West Bengal, India - 741235\\}
\begin{document}
\maketitle
\footnote{
Electronic address:\\
$^{1}$nayemsk1981@gmail.com\\}
\begin{abstract}

\noindent
Noether symmetry in teleparallel $f(T)$ gravity, where $T$ is the torsion scalar, has been studied in the background of Robertson-Walker space-time. It is found that Noether symmetry admits $f(T)\propto T^\frac{3}{2}$ and the associated conserved current is $\Sigma = a \dot a T^\frac{1}{2}$, in matter dominated era. In the process, the recent claim by Wei et.al. \cite{o} that Noether symmetry admits $f(T)\propto T^{n}$, (where $n$ is an arbitrary constant) is found not to be correct, since the conserved current satisfies the field equations only for a special choice of $n = {3\over 2}$. Further, correspondence between $f(R)$ and $f(T)$ theories of gravity has also been established.

\end{abstract}
Keywords:\\
$f(T)$ gravity, Canonical formulation, Noether symmetry.
\maketitle

\section{Introduction}

Luminosity-distance versus redshift curve obtained from distant SN1a supernovae data \cite{riess, perlmutter} unveils its non-linear feature beyond redshift $z = 0.2$. For last two decades attempts have been made to fit such a curve within a viable cosmological model. Almost all the attempts equivocally predict that at present the universe is undergoing an accelerated expansion. Since cosmological constant $(\Lambda)$ calculated in view of quantum field theory has been found to be nearly $120$ order of magnitudes larger than the same required to fit SN1a data, so it is ruled out. Two options therefore are left. The first is to modify the right hand side of Einstein's equation by accommodating one or more scalar fields including tachyonic fields or some more exotic ones having reverse sign in kinetic term with some typical form of potential. Such fields interact with none other than the gravitational field only, and therefore dubbed as dark energy. However, the field mass responsible for late time cosmic acceleration is very small on one hand and the present technology does not support detection of dark energy in any of its form, on the other. Therefore the second option has been advocated in recent years and that is to modify the left hand side of Einstein's equation, viz. the geometry, and in the process bypassing the dark energy issue. Such attempt is dubbed as modified theory of gravity. Several types of modified theory of gravity exists in the literature, such as, $f(R)$ gravity, $f(G)$ (Gauss-Bonnet) gravity, $f(T)$ (Torsion) gravity, combination of all these, Gauss-Bonnet-dilatonic coupled gravity, Lanczos-Lavlock gravity, Horava-Lifschitz gravity and models with extra dimensions including Kaluza-Klein, Randall-Sundrum, DGP and higher co-dimension braneworlds, etc. Out of these, Teleparallel gravity has drawn lot of attention in the recent years.\\

To consider teleparallelism, one employs the orthonormal tetrad components $e_{A} (x^{\alpha})$, where the index $A$ runs over 0, 1, 2, 3 to the tangent space at each point $x^{\alpha}$ of the manifold. Their relation to the metric $g_{\alpha\beta}$  is given by

\be\label{t1} g_{\alpha\beta}=\eta_{AB}e^{A}_{\alpha} e^{B}_{\beta},\ee
\noindent
where $\alpha$ and $\beta$ are coordinate indices on the manifold which again run over 0, 1, 2, 3, while $e^{A}_{\alpha}$ forms the tangent vector on the tangent space over which the metric $\eta_{AB}$ is defined. Instead of the torsionless Levi-Civita connection which is used in General Theory of Relativity, in Teleparallelism \cite{ben} one considers the curvatureless Weitzenbock connection, whose non-null torsion $T^{\rho}_{\alpha\beta}$ and contorsion $K^{\alpha\beta}_{\rho}$ are defined by

\be\label{t2} T^{\rho}_{\alpha\beta} \equiv e^{\rho}_{A}[\partial_{\alpha}e^{A}_{\beta}-\partial_{\beta}e^{A}_{\alpha}] ,\ee
\be\label{t3} K^{\alpha\beta}_{ \rho} \equiv  -\frac{1}{2}[{T^{\alpha\beta}}_{\rho}-{T^{\beta\alpha}}_ {\rho}-{T_{\rho}}^{\alpha\beta}],\ee
\noindent
respectively. Moreover, instead of the Ricci scalar $R$, which is used for the Lagrangian density in general relativity, the teleparallel Lagrangian density is represented by the torsion scalar $T$ given by
\be\label{t4} T \equiv {S_{\rho}}^{\alpha\beta} {T^{\rho}}_{\alpha\beta},\ee
where,
\be\label{t5} {S_{\rho}}^{\alpha\beta} \equiv  \frac{1}{2}[{K^{\alpha\beta}}_{\rho}+{\delta}^{\alpha}_{\rho}{T^{\theta\beta}}_ {\theta}-{\delta}^{\beta}_{\rho}{T^{\theta\alpha}}_{\theta}].\ee
\noindent
Accordingly, in analogy to the $f(R)$ theory of gravity, recently a new modified theory of gravity, namely the so-called $f(T)$ theory of gravity has been proposed to explain the current accelerated expansion of the cosmos, without invoking dark energy. Such a modified teleparallel action for $f(T)$ gravity is given by

\be\label{t6}\mathbb{ A} = \int  d^4 x  \mid e \mid  f(T)+ S_m ,\ee
where $|e|$ = det $e^{A}_{\alpha}=\sqrt {-g}$ and the units has been chosen so that $c = 16 \pi G = 1$. It may mentioned that this is a generalized version of the teleparallel gravity originally proposed almost a century back by Einstein \cite{3, 4}. \\

Now, in order to study the cosmological consequence of the so-called teleparallel gravity, a particular form of $f(T)$  is required. Instead of setting a form of $f(T)$ by hand or reconstruct it from the history of cosmic evolution, it is always desirable to find its form following some physical consideration, viz. in view of the loop quantum gravity or from some symmetry consideration. Since, loop quantum gravity does not provide a term suitable for late time cosmic acceleration, so Noether symmetry is usually preferred.\\

Noether symmetry was applied for the first time in scalar-tensor theory of gravity by De. Ritis and his collaborators \cite{ritis1} to find a form of the potential. Noether symmetry was found to select \cite{ritis1} exponential form of the potential which can trigger inflation in the early universe. This raise immense interest in the scientific community, and thereafter Noether symmetry has been extensively studied in cosmological models with minimally \cite{ritis2, rubano} and non-minimally coupled \cite{nmc, 22, 28, 31, 32} scalar-tensor theories, higher order theory \cite{23} and $f(R)$ theory \cite{c, v, sar1, ns4} of gravity. Additionally, the same has also been applied in different anisotropic Bianchi models \cite{30}, induced gravity theory \cite{33}, Gauss-Bonnet gravity \cite{35} and so on. Quantum cosmological models have also been expatiated in view of Noether symmetry \cite{29}. Here, we are therefore motivated to study Noether symmetry in teleparallel theory of gravity, to find a form of $f(T)$.\\

Recently, Wei et.al. \cite{o} has claimed that Noether symmetry for teleparallel $f(T)$ theory of gravity in the background of spatially flat Robertson-Walker (R-W) metric described by

\be\label{t6.1} {ds}^2 = - {dt}^2 + {a^2(t)} {dX}^2,\ee
where $a(t)$ is the scale factor, admits $f(T)\propto T^{n}$, where $n$ is an arbitrary constant in matter domain era. However, in the present study we show that the associated conserved current satisfies the field equations only for a special choice of $n = {3\over 2}$. Thus, it is found that Noether Symmetry only admits $f(T)\propto T^\frac{3}{2}$ along with a conserved current $\Sigma = a \dot a T^\frac{1}{2}$.\\

In the following section, the canonical formulations of $f(T)$ theory of gravity following Lagrange multiplier technique and its scalar-tensor counterpart have been discussed. In section 3, Noether symmetry has been invoked in both the canonical point Lagrangians corresponding to teleparallel $f(T)$ gravity. In section 4, analogy of teleparallel gravity with $f(R)$ theory of gravity has been discussed in some detail. Finally we conclude in section 5.

\section{Canonical formulation of f(T) gravity}

It is not possible to find solutions to the field equations corresponding to the above action \eqref{t6} to study cosmological consequence of teleparallel gravity, unless a specific form of $f(T)$ is known a priori. As already mentioned, one can choose a form by hand out of indefinite possibilities, or reconstruct it in view of cosmic evolution history. Nevertheless, it is always desirable to find the form in view of some physical consideration like Noether symmetry. Nevertheless, this requires canonical formulation of the theory under consideration. In fact, there exists two possible techniques towards canonical formulation of $f(T)$ theory of gravity. One is Lagrange multiplier technique, which is applicable with finite degrees of freedom, and the other is scalar-Tensor representation of $f(T)$ gravity.

\subsection{Lagrange multiplier technique}
Unlike Scalar-Tensor representation of $f(T)$ Theory (as we see next), canonical formulation following Lagrange multiplier technique may be performed with finite degrees of freedom only. Therefore we restricting ourselves to the Robertson-Walker metric (\ref{t6.1}), we can treat $T + 6 {\dot a^2\over a^2}=0$ as a constraint and introduce it in the action (\ref{t6}) through a Lagrange multiplier $\lambda$ as,

\bes\label{t7}\mathbb{ A}  = 2{\pi}^2\int\Big[f(T) - \lambda\Big\{T + 6\Big({\dot a^2\over a^2} \Big)\Big\}-\frac{\rho_{m0}}{a^3}\Big]a^3 dt .\end{split}\ee
Now varying the action with respect to $T$ one gets $\lambda = f'(T)$, where $f'(T)$ is the derivative of $f(T)$ with respect to $T$. Substituting the form of $\lambda$ so obtained in the above action \eqref{t7} the following canonical action is found, viz.
\bes\label{t8} \mathbb{ A} = 2{\pi}^2\int\Big[f(T) - f'(T)\Big\{T + 6\Big({\dot a^2\over a^2} \Big)\Big\}-\frac{\rho_{m0}}{a^3}\Big]a^3 dt .\end{split}\ee

\noindent
Therefore, the point Lagrangian in the presence of ordinary matter may be expressed in Robertson-Walker metric (\ref{t6.1}) as

\be\label{t9} L(a,\dot a,T,\dot T) = \left[-6a \dot a^2 f'+ a^3(f-f'T)-\rho_{m0} \right].\ee

\noindent
In the above, $\rho_{m0}$ stands for the matter density at the present epoch.

\subsection{Scalar-Tensor representation of f(T) gravity}
As already mentioned, it is also possible to translate the action (\ref{t6}) in its scalar-tensor equivalent form, in analogy to $f(R$) theory of gravity. The Scalar-Tensor representation \cite{ccl} of $f(T)$ gravity reads

\be\label{t13} \mathbb{ A} =  \int  d^4 x \mid h \mid[ \Phi T-U(\Phi)]+S_m.\ee
where,
\be\label{t14} \Phi =f'(T);~~~~ U(\Phi)= T f'(T) -f(T).\ee
The corresponding point Lagrangian in Robertson-walker (\ref{t6.1}) space-time reads

\be\label{t17} L(a, \Phi, \dot a, \dot \Phi) =\left[6a \dot a^2 \Phi  - a^3 U(\Phi)-\rho_{m0} \right],\ee
where, $\rho_{m0}$ is the matter density at the present epoch, as already mentioned.

\section{Noether symmetries }
In view of the canonical Lagrangians obtained in the previous subsections, we now move on to explore Noether symmetry. It is well known that Noether symmetry ($\pounds_X L = X L = 0$) in $f(R)$ theory of gravity yields nothing other than $f(R)=f_0 R^{\frac{3}{2}}$ along with a conserved current $\frac{d}{dt} (a\sqrt R)$ in R-W metric, when coupled to pressure-less dust or in vacuum \cite{c,v,sar1}. Despite such unique result, Noether symmetry of $f(R)$ theory of gravity had been reopened by some authors \cite{hus, jam}, who claimed to find new conserved currents in the name of Noether gauge symmetry. Particularly, it was claimed by Hussain et-al \cite{hus} that Noether gauge symmetry admits $f(R) \propto R^n$, where $n$ is an arbitrary constant. Jamil et-al \cite{jam} on the other hand found $f(R) \propto R^2$ and $V(\phi) \propto \phi^{-4}$, considering Noether gauge symmetry with Tachyonic field. The claim \cite{hus} had been reviewed by the present author and his collaborator (Sk and Sanyal) \cite{ns1} taking both vanishing and non-vanishing gauge into account. It was found that the conserved currents so obtained do not satisfy the field equations, particularly the ($^0_0$) equation of Einstein, unless $n = \frac{3}{2}$. Thus, the claim that arbitrary power of $R$ generates Noether symmetry is not correct. The claim of Jamil et-al \cite{jam} had also been reviewed by the same authors Sk and Sanyal \cite{ns2} and it was shown that $f(R) \propto R^2$ do not satisfy the Tachyonic field equations. Shamir et-al \cite{sam} on the contrary, had claimed that Noether symmetry of $f(R) \propto R^{\frac{3}{2}}$ admits four different generators corresponding to which four different conserved currents exist in the presence of non-zero gauge. In a subsequent communication, the same authors Sk and Sanyal \cite{ns3} reviewed the work and proved that the claim is not correct, for the same reason that not all the conserved currents satisfy the ($^0_0$) equation of Einstein. Later, Roshan et.al. \cite{rs} claimed that Noether symmetry in the context of Palatini $f(\Re)$ theory of gravity admits $f(\Re)\propto \Re^{n}$, (where $n$ is again an arbitrary constant) in matter dominated era. This claim had also been reviewed by the present author \cite{nsk} and it has been also shown that Noether Symmetry only admits $f(\Re)\propto\Re^\frac{3}{2}$ in Palatini gravity. Under such circumstances, it would really be interesting if $f(T)$ theory of gravity yields new forms of $f(T)$ as claimed by Wei et.al. \cite{o}. In the following subsections we therefore review the claim \cite{o} in the process of finding Noether symmetries of $f(T)$ theory of gravity, which satisfy the field equations.

\subsection{Noether symmetry following Lagrange multiplier technique}
The field equations constructed out of the point Lagrangian (\ref{t17}) in the Robertson-Walker metric (\ref{t6.1}) are,

\be\label{t17.1} \left(f-f'T+2f'H^2 \right) + 4\left( 2f' \frac{\ddot a}{a}+H f''\dot T\right)=0 ,\ee

\be\label{t17.2} a^3 f''\left(T+6\frac {\dot a^2}{a^2} \right)=0 .\ee
In the above $H = {\dot a\over a}$ stands for the Hubble parameter. The ($^0_0$) equation of Einstein is

\be\label{t17.3}\left[-6a \dot a^2 f'+ a^3(f-f'T)-\rho_{m0} \right]=0 .\ee
Now, Noether theorem state that, if there exists a vector field $X$, for which the Lie derivative of a given Lagrangian $L$ vanishes i.e. $\pounds_X L = X L = 0$, the Lagrangian admits a Symmetry and thus yields a conserved current. For the Lagrangian \eqref{t9} under consideration, the configuration space is $M(a,T)$ and the corresponding tangent space is $\mathrm{T}M(a,T,\dot a, \dot T)$. Hence the generic infinitesimal generator of the Noether Symmetry is

\be X = \gamma \frac{\partial }{\partial a}+\zeta\frac{\partial }{\partial T} +\dot\gamma \frac{\partial }{\partial\dot a}+ \dot\zeta\frac{\partial}{\partial\dot T},\ee
where, $\gamma = \gamma(a, T), \zeta = \zeta(a, T)$. The constant of motion is given by
  \be \Sigma = \gamma \frac{\partial L }{\partial\dot a}+ \zeta\frac{\partial L }{\partial\dot T} .\ee
Finding the Noether equation in view of the existence condition $\pounds_X L= X L = 0 $, and equating the coefficients of $\dot{a}^2$, $\dot{T}^2$, $\dot a \dot T$ along with the term free from derivative respectively to zero as usual, we obtain the following set of partial differential equations,
\bes\label{t18}
   a\gamma' = 0, \;\;
   \gamma f' +  \zeta a f''+ 2a f'\gamma_{,a}= 0, \;\; \\&
   3\gamma \left({f-T f'}\right)-a\zeta T f'' =0.
\end{split}\end{equation}
The above set of partial differential equations admit the following set of solutions, viz.
\be\gamma = \gamma_0 {a}^{1-s} ,\;\;\;\;\zeta= -2 s\gamma_0 {a}^{-s} T,\;\;\;\; f(T) = f_0 T^{\frac{3}{2s}}.\ee
The corresponding conserved current is
\be\label{con} \Sigma=  a^{2-s}\dot a f'(T).\ee
It may be trivially checked that the above conserved current satisfies the field equations \eqref{t17.1} through \eqref{t17.3} only for $s = 1$. The expression of the conserved current \eqref{con} for $s =1$ therefore reads,
\be \label{cons}\Sigma=  a\dot a f'(T).\ee
It is interesting to note that the reduced form of $f(T)$ turns out to be,
\be f(T)=f_{0}T^\frac{3}{2}.\ee
\noindent
In view of the above form of $f(T)$ and the conserved current \eqref{cons}, $\dot a(t)$ turns out to be a constant, and therefore the cosmic scale factor $a(t)$ admits the solution,
\be a(t)= a_1 t + a_0,\ee
where $a_1, a_0$ are constants of integration. In this context we mention that the same solution \cite{nsk} has also been found in the context of Palatini $f(\Re)$ theory of gravity. However, the above coasting solution although fits SnIa data perfectly in the matter dominated era \cite{sar2}, does not fit to other available cosmological data.

\subsubsection{Comments on  Hao Wei et al. work }

It is important to note that the ($^0_0$) equation of Einstein is essentially the Hamiltonian constraint equation, when expressed in terms of phase-space variables. It is the outcome of diffeomorphic invariance of the theory of gravity. Since Noether equation $\pounds_X L = X L = 0$ does not recognise the constraint, therefore one can not expect that the solutions of Noether equations would satisfy the Hamilton constraint equation automatically. This has been proved by Wald and Zoupas \cite{wz}. This means that Noether theorem is not on-shell for constrained system. Conserved current is not an independent equation, but rather it is the first integral of certain combination of the field equations. Thus, it is essential to check if the conserved current satisfies the ($^0_0$) equation of Einstein. Like earlier authors \cite{hus, jam, sam, rs} it has not been checked by the present authors \cite{o}. However, it is not difficult to check that the conserved current satisfies the field equations only under the special choice $n=\frac{3}{2}$. Therefore, the claim of finding $f(T) \propto T^n$ by Wei et al. \cite{o} is not correct.

\subsection{Noether symmetry in Scalar-Tensor representation of f(T) gravity}

Let us now turn our attention in this subsection, to explore Noether symmetry in scalar-tensor representation of $f(T)$ theory of gravity. The field equations constructed out of the point Lagrangian (\ref{t13}) in the Robertson-Walker metric (\ref{t6.1}) are,
\be\label{t18.1}\left[\frac{\ddot a}{a} + \frac{\dot a^2}{2a^2} + \frac{\dot a\dot \Phi}{a \Phi} + \frac{U}{4\Phi}\right] =0,\ee

\be\label{t18.2}\left[ \frac{\dot a^2}{a^2} - \frac{U_{,\Phi}}{6}\right] =0.\ee
The ($^0_0$) equation of Einstein is

\be\label{t18.3}\left[6a \dot a^2 \Phi+ a^3 U(\Phi)+\rho_{m0} \right]=0 .\ee
In order to apply Noether symmetry approach, let us again introduce the lift vector $X$ as an infinitesimal generator of Noether symmetry in the tangent space $[a,\dot a,\Phi,\dot \Phi]$ as follows

\be\label{t19} X = \gamma \frac{\partial }{\partial a}+\zeta\frac{\partial }{\partial \Phi}  +\dot\gamma \frac{\partial }{\partial\dot a}+ \dot\zeta\frac{\partial }{\partial\dot \Phi} , \ee
and the existence condition for symmetry, $ X L = 0 $, leads to the following system of partial differential equations
\bes\label{1.91}
   \gamma_{,\Phi} =0 ,~~~  \Phi\gamma+a\zeta +2a\Phi\gamma_{,a} = 0,~~~ 3\gamma U +a\zeta U_{,\Phi} =0.
\end{split}\ee

\noindent
The solution of the above set of equations reads,
\be\label{t21}
    \gamma=-\gamma_0 d{a}^{\frac{1-d}{2d}},~~~ \zeta=\gamma_0 {a}^{\frac{1-3d}{2d}}{\Phi},~~~ \ U=U_0{\Phi}^{3d},\ee
\noindent
while the expression of conserved current is
\be\label{t22}  \Sigma=  a^\frac{1+d}{2d}\dot a \Phi.\ee

\noindent
Again, it has been shown that the above conserved current satisfies the field equations (25) to (27) only for $d=1$. The expression for the conserved current for $d =1$ is therefore,
\be \Sigma=  a\dot a \Phi.\ee
\noindent
Now, using the transformation relations \eqref{t14}, we rewrite $U(\Phi)$ as $U(\Phi) = T f'(T)- f(T) = U_{0}\Phi^3 = U_{0} [f'(T)]^3$. Equation (\ref{t21}) therefore yields the following form of $f(T)$, viz.
\be\label{t23} f(T) = f_{0} T^{\frac{3}{2}},\ee
Note that the form of $f(T)$ and the associated conserved current so obtained is identical with those obtained following Lagrange multiplier technique. The solution to the scale factor therefore remains unchanged

\be\label{t24} a(t) = a_1 t + a_0,\ee
which as already stated is not a viable solution to fit available cosmological data. Nevertheless, one important issue has been revealed and that is Noether symmetry is independent on the choice of the configuration space variables.\\

\noindent
One of the main advantages of Noether conserved current is that one can express the field equations in terms of a cyclic coordinate, so that finding solutions becomes easier, and sometimes the cosmological solution emerges directly from Noether conserved current \cite{rnc}. Being a first integral, one can even use it to find the solutions without even finding cyclic coordinate. In any case, one has to use the conserved current to find the solutions to the field equations. In a recent article \cite{bcl}, power-law teleparallel $f(T)$ gravity is discussed in details. The authors first applied Noether symmetry to find the form $f(T)\propto f_{0}T^n$, and the associated conserved current. Thereafter, they explored the cosmological solution of the above mentioned form of $f(T)$ analytically with the help of the field equations and claimed the solutions to be outcome of Noether symmetry. One can easily check that the solutions do not satisfy Noether conserved current. Therefore such solutions cannot be an outcome of Noether symmetry, rather, it is like setting a form of $f(T) \propto f_{0}T^n$ by hand, and solving the field equations.

\section{Analogy with $f(R)$ gravity}

The teleparallel $f(T)$ gravity is not equivalent to metric $f(R)$ gravity in general, since they differ by an appropriate boundary term $(\mathbb{B})$ \cite{bbw,bs}. The relation  between torsion scalar $(T)$ and the Ricci scalar $(R)$ is given by

\be\label{t26} R = -T +\frac{2}{e}\partial_{\rho}(e T^{\rho})= -T+\mathbb{B},\ee
where, $\mathbb{B}=\frac{2}{e}\partial_{\rho}(e T^{\rho})$ is the boundary term. The action generally can be expressed as
\be\label{t26.1}\mathbb{A_{B,T}} = \int  d^4 x  \mid e \mid  f(T,\mathbb{B})+ S_m .\ee
It has been mentioned in a recent article \cite{bs} that both the metric $f(R)$ and the teleparallel $f(T)$ gravity can be recovered from $f(T,\mathbb{B})$ theory of gravity under suitable limit. Now, the expressions of torsion scalar $(T)$ and boundary term $(\mathbb{B})$ in a flat R-W metric are, $T = -6({\dot a^2\over a^2})$ and $\mathbb{B}=-6({\ddot a\over a}+2{\dot a^2\over a^2})$. Therefore, the Ricci scalar is
\be\label{t27} R = -T+ \mathbb{B}=-6({\ddot a\over a}+{\dot a^2\over a^2}).\ee\\
In this present article, we observe that Noether symmetry of teleparallel $f(T)$ theory of gravity in matter dominated only yields $f(T) \propto f_{0} T^{\frac{3}{2}}$. Such a form of $f(T)$ admits a solution of the cosmological scale factor, $a(t) = a_1 t + a_0$, in a flat R-W metric. This particular solution implies $\ddot a=0$. On the contrary, $f(R) \propto R^{3\over 2}$ yields a cosmological solution $a(t) = \sqrt{a_4 t^4 + a_3 t^3 + a_2 t^2 + a_1 t+ a_0}$. So, in general the two differs. However, when $t$ is small enough, i.e. in the early matter dominated era, the two match. In particular, under the condition $\ddot a=0$, teleparallel $f(T)$  gravity becomes equivalent to metric $f(R)$ gravity, since  $\mathbb{B}= 2T$ and  $R=T$, in view of equation (37). Essentially, Noether symmetry puts up a limit under which the two theories become equivalent. This clearly demonstrates that at least in the context of Noether symmetry it is practically of no use to consider teleparallel gravity over $f(R)$ theory of gravity.

\section{Concluding remarks}

In the present work we studied teleparallel gravity and explored the form of $f(T)$ invoking Noether symmetry in the background of isotropic and homogeneous R-W metric. Both the canonical point Lagrangians obtained following Lagrange multiplier method and the scalar-tensor equivalent one, have been found to admit the only symmetry $f(T) = f_{0} T^{\frac{3}{2}}$ in the matter dominated era. This reveals the fact that Noether symmetry, when applied to explore the form of an unknown parameter, is independent of the choice of the configuration space variables. We have also noticed that in R-W metric, Noether symmetry yields identical form of the cosmic scale factor $(a(t) = a_1 t + a_0)$ both in teleparallel $f(T)$ theory of gravity and Palatini $f(\Re)$ theory of gravity \cite{nsk}, in the matter dominated era. This establishes a sort of equivalence between the two. It has also been demonstrated that in the context of Noether symmetry teleparallel gravity turns out to be a special case of $f(R)$ theory of gravity.\\

It is clear that the form of $f(T)$ so obtained is not much appreciable. This is because, the coasting solution so obtained although fits SnIa data perfectly in the matter dominated era \cite{sar2} fails to fit other available cosmological data. Particularly, it does not admit a long Friedmann-like matter dominated era, prior to the recent accelerated expansion of the universe. So application of Noether symmetry to choose a form of $f(T)$ becomes useless. In this context, we would like to mention that recently it has been observed that indeed Noether symmetry of $f(R)$ theory of gravity yields forms other than $f(R) \propto R^{3\over 2}$ \cite{ss}. In particular the other forms are $f(R) \propto R^2, {1\over R}, {R^{7\over 5}}$. However, this requires a new symmetry generator, which includes the ($^0_0$) equation of Einstein in the form $\pounds_X L - \eta H = X L-\eta H =0$, where, $H$ is the Hamiltonian constraint of the theory being expressed in terms of configuration space variables (the $^0_0$ equation of Einstein) and $\eta$ is a function of the coordinates in general. There is also possibility of finding other forms of $f(R)$ under proper investigation. Likewise, we do expect that several other forms of $f(T)$ might also emerge in view of the above mentioned symmetry generator. This we pose in a future communication.\\

\noindent
\textbf{Acknowledgement}: I would like to thank my supervisor Dr. A. K. Sanyal for useful discussions and language editing. I also thank to reviewer's for their comments to improve this  manuscript .

\end{document}